# Starburst Galaxies

**T.W.B. Muxlow[1]**
*MERLIN-VLBI National Facility, Jodrell Bank Observatory,*
*Lower Withington, MACCLESFIELD, Cheshire, SK11 9DL, U.K.*
*E-mail:* twbm@jb.man.ac.uk

**R.J. Beswick, A.M.S. Richards, H.J. Thrall,**
*Jodrell Bank Observatory,*
*Lower Withington, MACCLESFIELD, Cheshire, SK11 9DL, U.K.*
*E-mail:* rbeswick@jb.man.ac.uk, amsr@jb.man.ac.uk, hthrall@jb.man.ac.uk

Star-formation and the Starburst phenomenon are presented with respect to a number of nearby star-forming galaxies where our understanding of the process can be calibrated. Methods of estimating star-formation rates are discussed together with the role played in the investigation of the process by multi-wavelength studies of a few selected starburst galaxies (especially the well studied galaxy M82). Our understanding of nearby systems allows us to study the star-formation history of the Universe by observing high-redshift starburst galaxies. These begin to dominate the radio source populations at centimetric wavelengths at flux densities below a few 10s of μJy. New very sensitive, high resolution telescopes in the sub-mm and radio will revolutionize our understanding of these distant star-forming systems, some of which may contain embedded AGN.

---

[1] Tom Muxlow

## 1. What is a Starburst Galaxy?

Starburst galaxies are characterised by high star-formation rates of order 10 – 100 M☉/year. These are much higher than are normally found in gas-rich normal galaxies (eg Milky Way star-formation rate ~ 1 – 5 M☉/year). The total gas content of a galaxy can be estimated from integrated HI line profiles from which we can derive the HI mass. From the gas available to fuel the star-formation event and the observed star-formation rate we can derive the sustainable lifetime of the star-formation event. These are typically a few x $10^9$ years for objects like the Milky Way – which means that the present level of star-formation can be maintained for the lifetime of the galaxy (~$10^8$ years). However, for a starburst galaxy, the lifetimes are short compared with the galaxy age implying that there is a 'burst' of star-formation which can only be sustained for a relatively short period on the cosmic timescale.

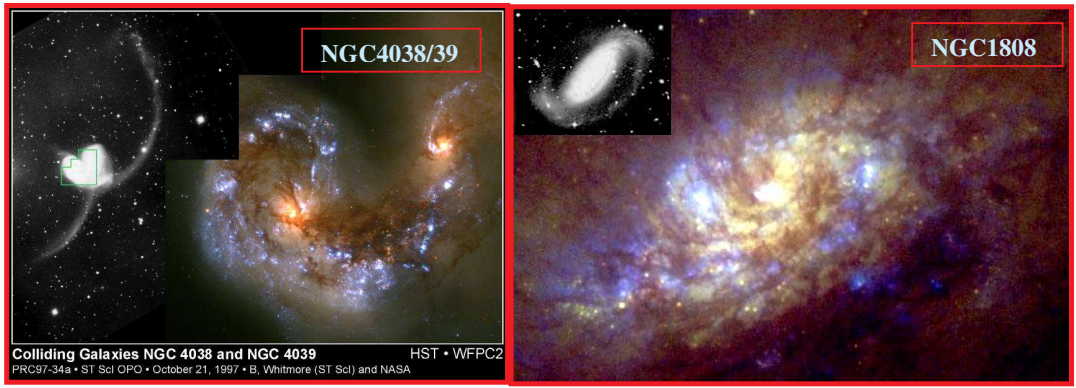

*Figure 1: HST images of the central star-forming regions in the starburst galaxies NGC4038/39 & NGC1808. HST image archive & Nobeyama radio observatory.*

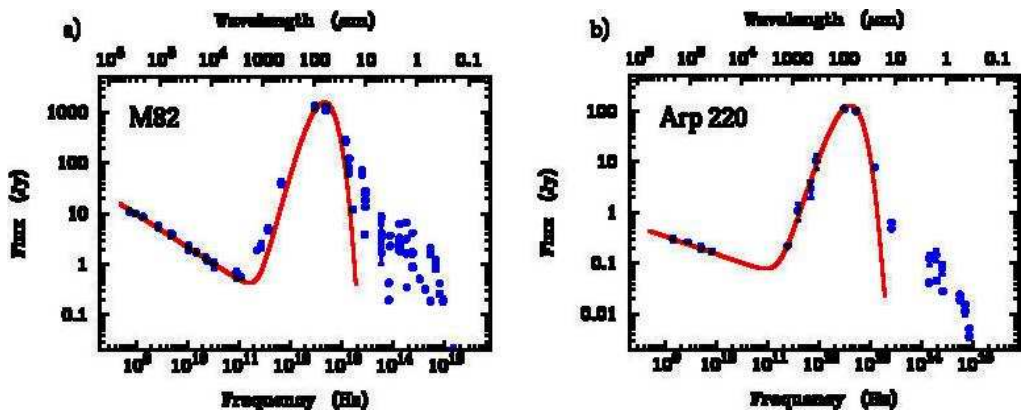

*Figure 2: Spectral Energy Distribution of the nearby star-forming galaxies M82 and Arp 220 .*

Typically starbursts are disturbed galaxies, and many are observed to be merging systems. Star-formation sites within the galaxy are usually marked out by statistically rare highly-luminous high-mass blue stars. The spectral energy distribution (SED) is often dominated by strong infra-red (IR) and far infra-red (FIR) emission, (See figure 2).

The size of starburst regions is typically a few kpc in extent. Non-thermal radio synchrotron emission is detected from the hot plasma ejected from recent supernovae and strong FIR emission is found from UV-heated dust (from O stars). FIR luminosities are high, with $L_{FIR} > 10^{10} L_\odot$ (Ultra-Luminous Infra-Red Galaxies (ULIRGs) have $L_{FIR} > 10^{12} L_\odot$).

## 1.1 Star-formation Rate Indicators

Star-formation rate (SFR) indicators are many and varied. These have been extensively studied - eg [1]. There are a number of possible indicators, some of which are better than others. Some of the best are the FIR (60μm) and radio continuum (1.4GHz) luminosities since these are essentially extinction-free.

The SFR estimated from FIR and 1.4GHz radio continuum are found to be highly correlated over many orders of magnitude. The SFR can also be estimated from the numbers of O stars required to ionize the inter-stellar medium and produce thermal free-free continuum emission and forbidden lines (eg [NeII]). This can be probed by measuring the UV flux directly, however such observations suffer from extinction due to the large quantities of gas and dust situated within the centres of starburst galaxies. Measuring thermal radio free-free emission does not suffer from these extinction problems, but it is difficult to separate from the steep-spectrum non-thermal radio emission. Alternatively, the supernova rate can be used to derive the SFR for stars more massive than 8M$_\odot$ however dust extinction may lead to an underestimate of the true supernova rate [2].

## 2 Nearby Starburst Galaxies

Amongst the best studied nearby star-forming galaxies are M82 (3.2Mpc) and Arp220 (77Mpc).

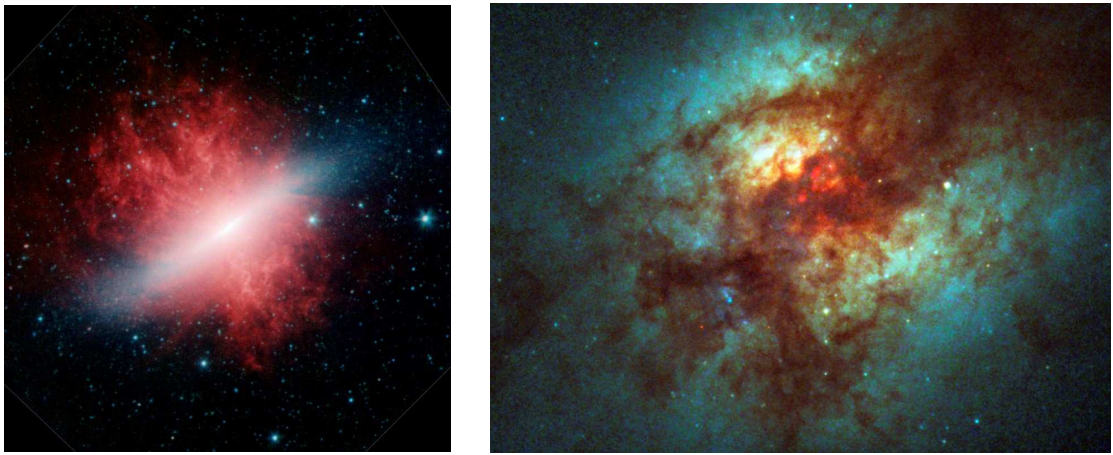

*Figure 3: Spitzer IR image of M82(left) showing outflow from the central starburst and (right) HST image of the central double nucleus region of Arp 220. Credit NASA.*

### 2.1 Arp220

This is a merging ULIRG system with two nuclei. HI is detected in emission with the VLA in C+D array at low angular resolution. However, towards the central starburst the HI distribution is dominated by absorption [3] as shown in figure 4. HI absorption studies at high (sub-arcsecond) angular resolution with MERLIN have been used to probe the merger dynamics [4]; where two counter-rotating disks have been found – these are the original galaxy cores that have survived the initial encounter and are now in the final stages of merger.

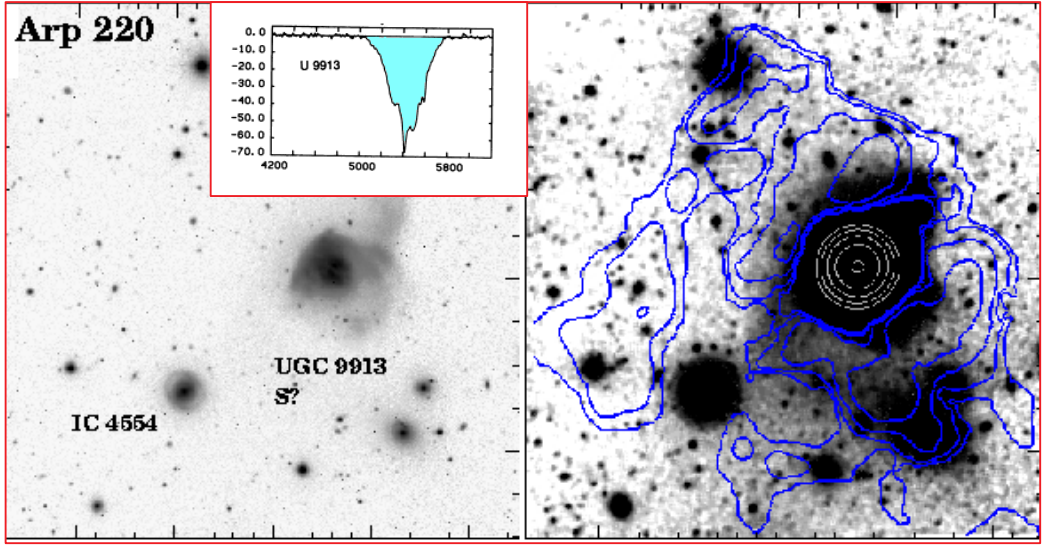

*Figure 4: VLA C+D array image of ARP 220 with an angular resolution of 30" shown as contours overlaid on a deep R-band The central double nucleus is dominated by absorption . [3]*

OH mega-masers within the eastern disk show a velocity gradient of 320 km s$^{-1}$ kpc$^{-1}$. With a radius of ~80pc this implies an enclosed mass ~$1.2 \times 10^7$ M$\odot$. Thus there is no $10^8$ M$\odot$ black hole, but the Chandra satellite has detected hard X-rays from the western disk suggesting AGN activity.

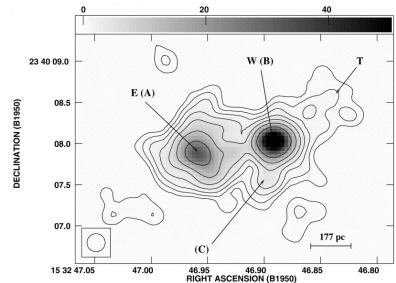

*Figure 5: MERLIN image of the central double nucleus region of Arp 220. [4]*

Global VLBI images with an angular resolution of a few mas show that both the east and west components contain radio Sn and SNR. There is no direct evidence for an active AGN seen, but in the western nucleus, unusual structures &

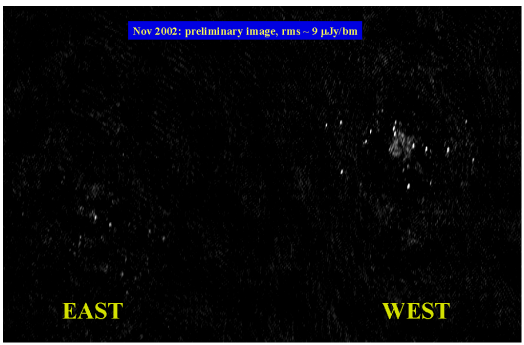

*Figure 6: Global VLBI image of the central double nucleus region of Arp 220. [5]*

velocity gradients are detected [5].

## 2.2 M82

M82 is one of the nearest (distance=3.2 Mpc) & best studied starburst galaxies. The starburst located in nuclear region with a size of ~1 kpc. Spectacular starburst-driven winds have detected in both Hα and X-ray emission, as can be seen in a recent Subaru telescope image – see figure 7. The starburst activity has been triggered by a tidal interaction over the past ~200 Million years with the nearby M81/NGC3077 system. M82 is in high-speed motion with respect to the M81/NGC3077 group as shown by HI observations [6]. Several 'super' star-clusters have been identified ~1kpc to the north-east of the centre of M82. Their ages have been derived from evolutionary spectral synthesis models and indicate that these clusters have ages of around $10^{9.}$ years implying that the starburst activity may be episodic on such a timescale - which is close to the epoch of the previous encounter of M82 with M81 [7].

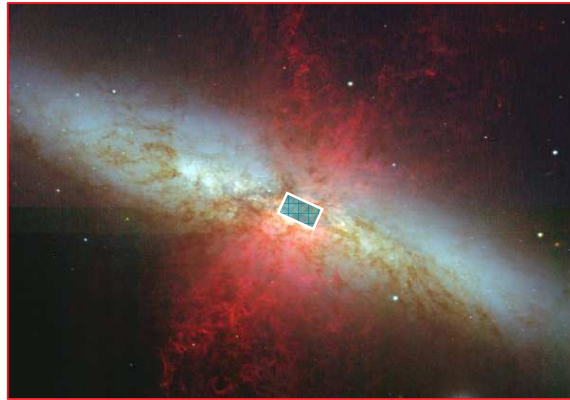

*Figure 7: Hα filaments showing a starburst-driven outflow. The starburst region is confined to the central ~1kpc of the galaxy and the extent of the radio image shown in Figure 8 is shown as a box. - Credit Subaru telescope*

HI absorption has been detected across the face of the central starburst region by the VLA [8]. The velocity field shows strong rotation and a detailed fit along major axis suggests the presence of an inner bar which is thought to be fuelling the central starburst by channelling gas into the central nuclear region. This is also seen in molecular lines (eg CO) [9].

O & B type stars are tracers of recent star-formation. Typically these high-mass stars become supernovae after ~ $10^7$ years, thus SNR trace out star-formation sites which are typically ~$10^7$ years old. Between 50 and 60 compact sources have been discovered in M82. All are resolved with both MERLIN and VLBI. Most are thought to be SNR - although ~16 are have rising thermal spectra and are identified at compact HII regions. Since MERLIN resolves all the SNR visible in M82 it is possible to derive the SNR size distribution [12]. The cumulative number-size diagram (Figure 9) infers that SNR expansion slows with time: $D \sim t^{0.6}$ until the average SNR surface brightness matches that of the extended background, when they then become very difficult to detect. High resolution radio observations over several epochs have allowed the expansion velocities of the smaller remnants to be measured directly [13],[14]. Assuming such velocities are typical, the compact SNR in M82 are all younger than ~2000 years.

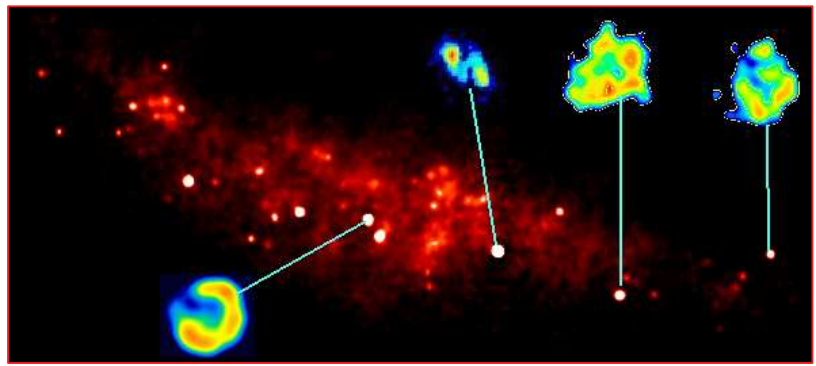

*Figure 8: MERLIN+VLA image of M82 at 5GHz. Insert images are MERLIN and VLBI images of individual SNR at high angular resolution.*

## 2.3 VLBI Imaging of SNR

This has been performed on M82 and other nearby starburst galaxies (See figure 10). This has enabled workers to measure the expansion velocities of many individual SNR and to study the deceleration of the ejected material.

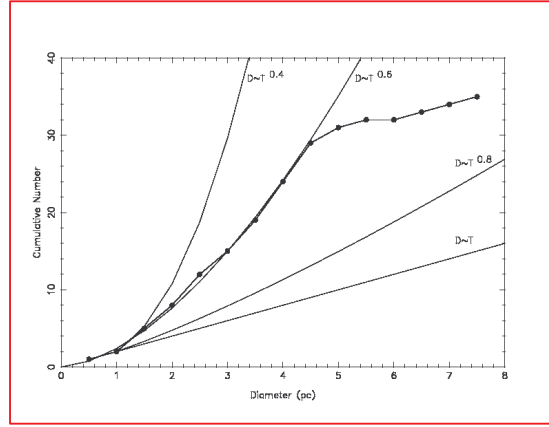

*Figure 9: Cumulative-size distribution for compact SNR in a 5GHz MERLIN image of M82.*

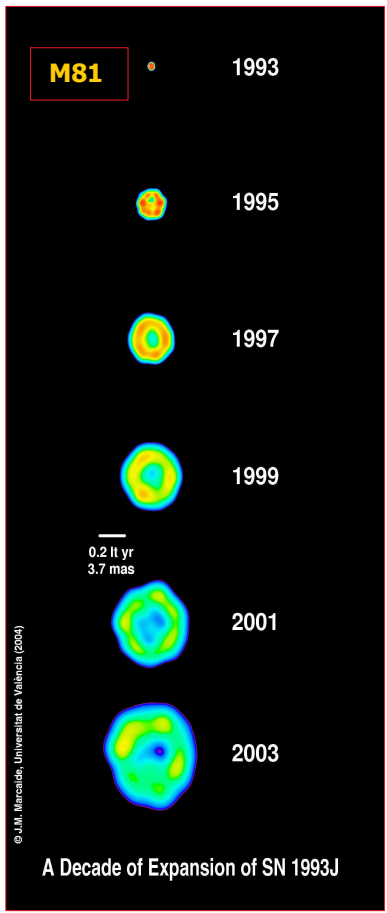

It is thus possible to probe the nature of the environment, which is thought to be extremely clumpy in many cases; and to investigate how the ejecta interact with the SNR environment and eventually move into the Sedov phase of expansion where the swept-up mass begins to be comparable with that that ejected. A number of important questions still remain unanswered – for example: Do supernovae in environmental voids produce no observable radio remnant? In M82, the SFR as derived from the 1.4GHz radio and FIR luminosities implies a SN rate of 1 every ~12 years. By comparison, the observed SNR rate is closer to 1 every ~30 years.

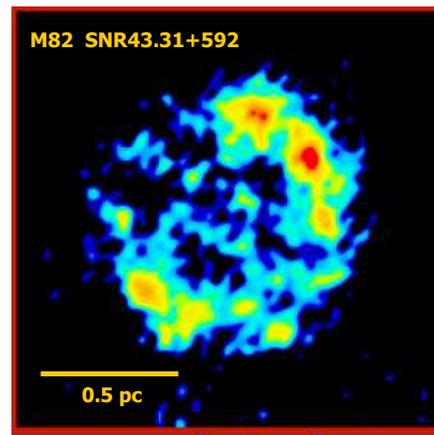

*Figure 10: (Left) VLBI images of SNR 1993J in M81 [15],[16] and (Right) SNR 43.31+592 in M82.*

## 3  Embedded AGN

Some starburst galaxies have been found to contain embedded AGN. Often the AGN is obscured and may only contribute a small proportion of the total flux density of the source. One example is the ULIRG Mrk273 which contains twin merging nuclei [17]. MERLIN high resolution HI absorption studies [18] found the dynamical signature of a super-massive object in a rotating disc with a mass of ~ $1.4 \times 10^{10}$ M☉ in a region smaller than ~500pc. VLBA and EVN observations [19],[20] have found a compact AGN candidate coincident with one of the two nuclei containing ~$2 \times 10^9$ M☉ in a region smaller than 220pc (Figure 11). The object could be a compact radio supernova, but it is found to emit hard X-rays [21] which suggests that it is an AGN.

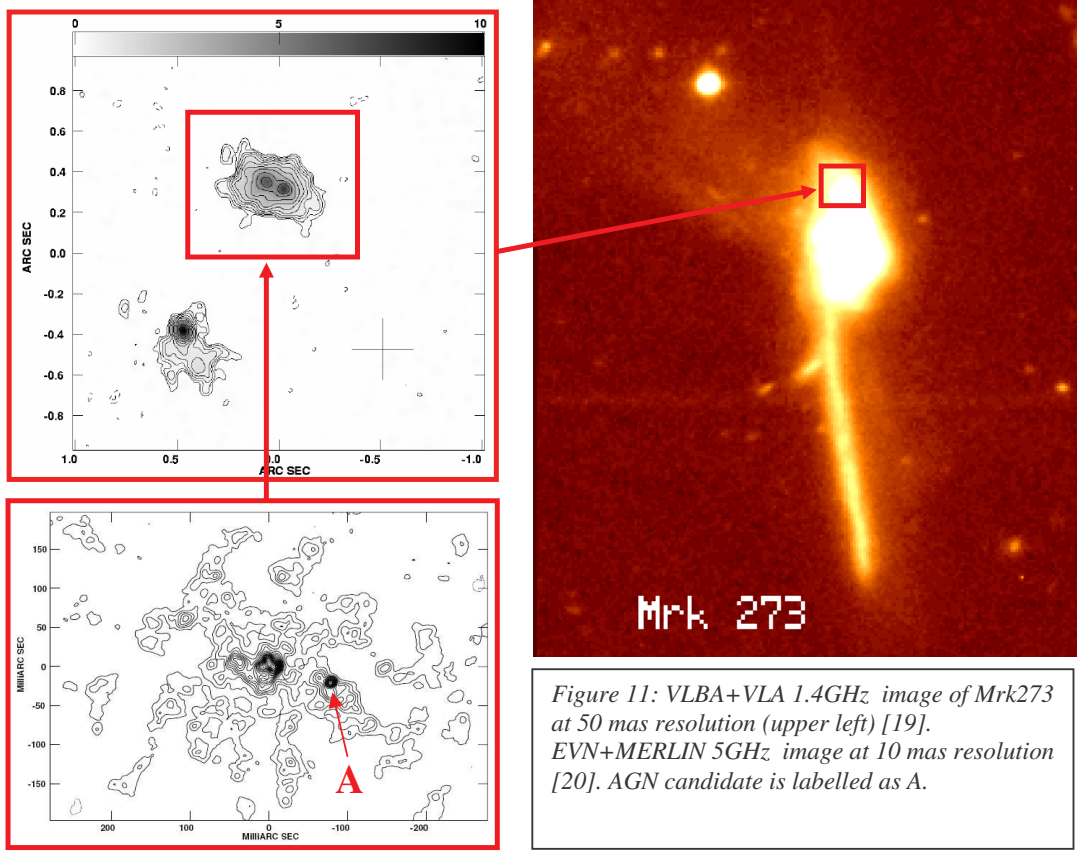

*Figure 11: VLBA+VLA 1.4GHz image of Mrk273 at 50 mas resolution (upper left) [19]. EVN+MERLIN 5GHz image at 10 mas resolution [20]. AGN candidate is labelled as A.*

## 4. Starformation at High Redshift

Deep galaxy studies indicate that early galaxies merge to form larger systems in a 'bottom-up' scenario of galaxy assembly. This implies that galaxy-galaxy interactions were common at early epochs. Such interactions are likely to trigger major star-formation activity. Multi-wavelength studies of a number of fields including deep radio observations involving VLA, ATCA, MERLIN, EVN…[21],[22],[23] have shown that at flux densities <1mJy at 1.4GHz there is new population of faint radio sources that are associated with distant star-forming galaxies.

### 4.1 Deep HDF-N MERLIN + VLA High Resolution Imaging

One of the most intensively studied regions in the northern sky is the Hubble Deep-Field North (HDF-N). Low-resolution VLA imaging of this field [21],[22] has identified a population of µJy radio sources. The inner central 10 arcmin$^2$, centred on the HDF-N has been imaged at high angular resolution (0.2 – 0.5 arcseconds) with a combination of MERLIN and the VLA [24]. This deep combination imaging has produced an image with an rms noise of ~3.3µJybm$^{-1}$ - one of the most sensitive 1.4GHz images yet made, and has revealed the radio structures of this µJy source population for the first time.

Results from the original study of the 10 arcminute$^2$ field has shown:

- 92 radio sources with flux densities >40µJy at 1.4GHz
- Angular sizes in the range 0.2"–3"
- 85% of the sources are associated with galaxies brighter than 25th mag in V-band
- The remaining 15% are optically faint EROs at high redshift (some are detected by SCUBA as sub-mm sources)
- Below ~60µJy sources are dominated by starburst systems (See figure 12)
- Some luminous high-redshift starbursts show evidence for an embedded AGN

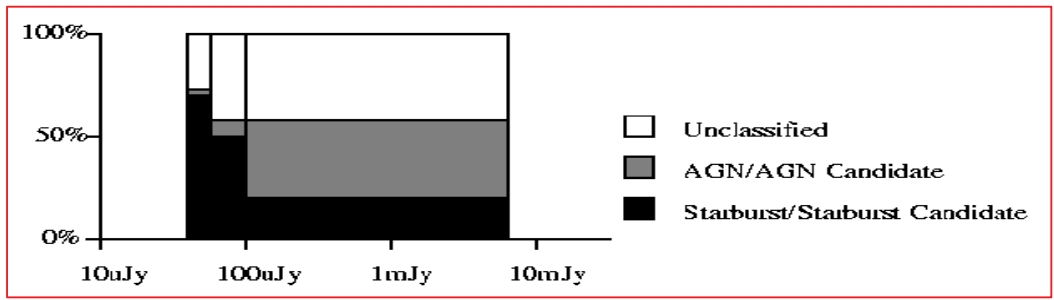

*Figure 12: The proportion of starburst systems as a function of radio flux density at 1.4GHz from [24]. Below 60µJy >70% of the radio sources are starburst systems [24].*

Starburst systems were identified as radio sources with steep radio spectral index and with radio structures extended on (sub-)galactic scale sizes overlaying the central region of the optical galaxy. Those starburst systems containing an embedded AGN have compact VLBI cores [25],[26] (See figure 13), and/or hard X-ray emission.

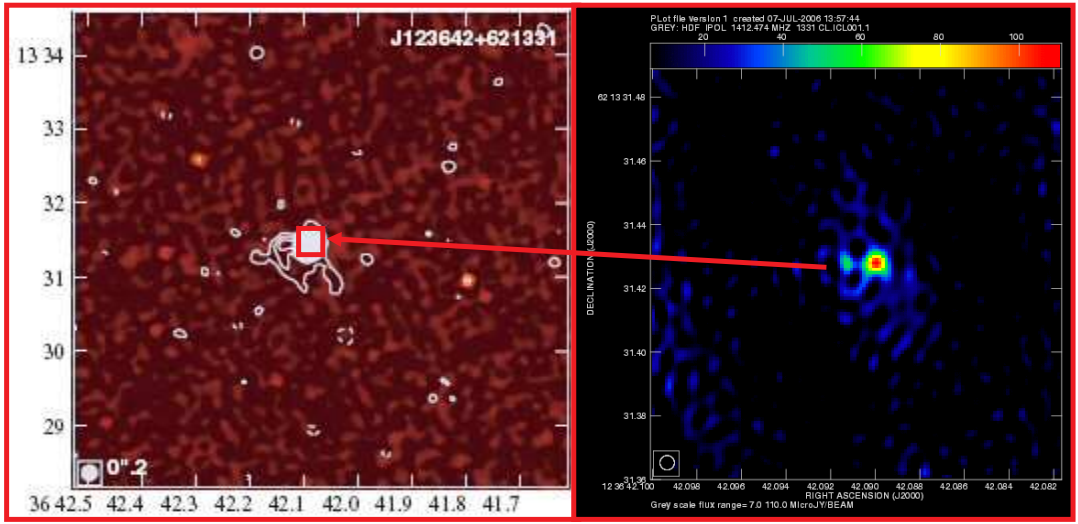

*Figure 13: VLBI image of an embedded AGN core (right in false colour) within the high-redshift (z=4.424) starburst system J123642+621331. The MERLIN/VLA combination map is shown contoured (left) overlaid on the HST WFPC2 v-band image [24],[25].*

### 4.2 Extension of the Original Study to Include GOODS North

New HST ACS & Spitzer data are now available for the GOODS field North. The historical HDF-N is coincident with the GOODS north field, however the GOODS region of study extends over a significantly wider area than the original HDF-N deep WFPC2 images.

An 8.5' x 8.5' MERLIN+VLA radio field centred on HDF-N was found to intersect with 13030 galaxies brighter that 28.3mag in ACS z-band field from the GOODS North field. Using these new data we have now extended the initial analysis presented in [24], to investigate statistically the very weak radio source population below with flux densities at 1.4GHz of less than 20μJy.

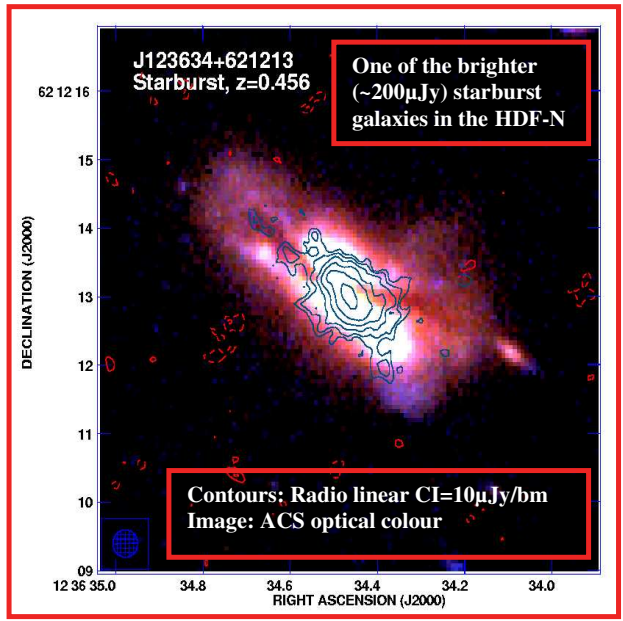

*Figure 14: MERLIN/VLA radio image of starburst system J123634+621213 overlaid on HST ACS image.*

### 4.3 Radio Emission From ACS Galaxies:

The radio flux density from each of the 13030 z-band optical galaxies was investigated by integrating the radio emission from the MERLIN+VLA combination image of the 8.5 arcminute field within a radius of 0.75 arcsecond at the position of every galaxy. Radio sources brighter than 40µJy had already been studied in detail in [24] and were excluded from the sample. In addition, all galaxies with nearest neighbours closer than 1.5 arcseconds were also excluded so as to avoid confusion problems. The radio and optical fields are astrometrically aligned to better than 50mas. The median flux densities, binned by z-band magnitude are shown in figure 15. The control sample was constructed which incorporated a random 7 arcsecond positional shift between the optical galaxy as the position searched in the radio image.

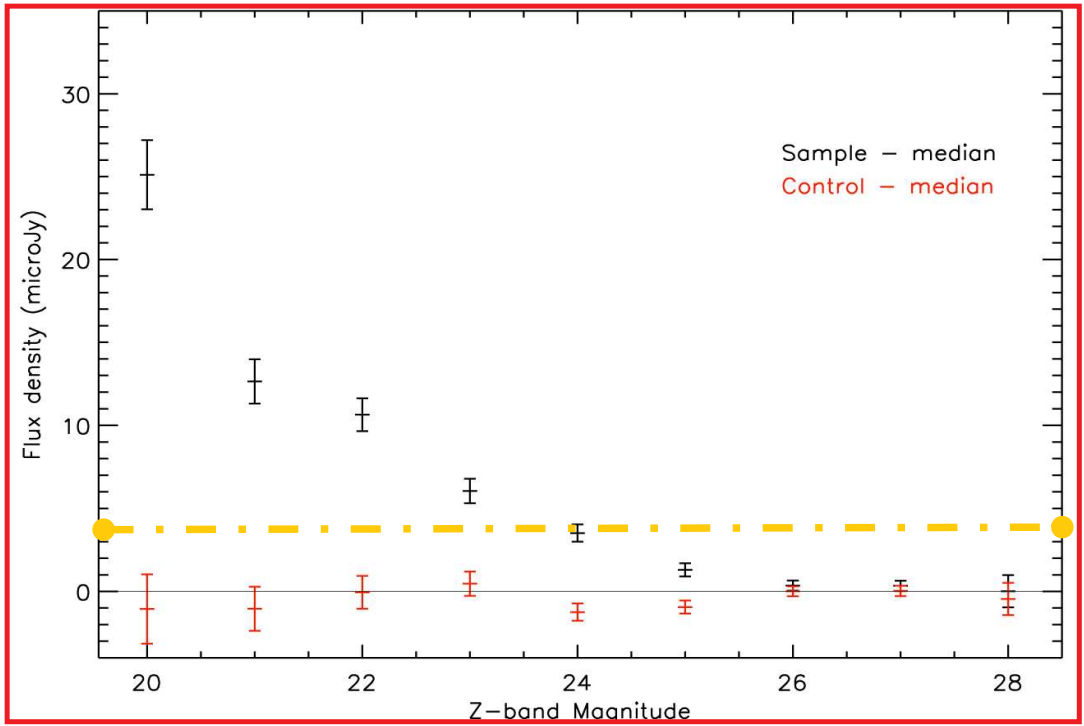

*Figure 15: The median 1.4GHz flux density detected within 0.75'' of the position of optical galaxies in the ACS HDF-N field, binned by z-band magnitude. Control sample incorporates a random 7" shift.*

Radio emission is clearly detected statistically for all magnitude bins brighter than a Z-band magnitude of 26. Of the ~2700 galaxies brighter than $Z=24^{mag}$, half (nearly 1400) will have radio flux densities of ~4µJy or greater at 1.4GHz. 4µJy represents an ~8σ detection for a future deep *e*-MERLIN/EVLA image; such galaxies which at present may only be investigated statistically will in the future be imaged individually.

Radio source sizes for these very weak radio sources can also be investigated statistically. The average radio source sizes in each magnitude bin can be derived from the radio flux densities found in annuli of width 0.5 arcseconds, over radii of 0.25 to 2 arcseconds centred on each galaxy position. These are shown in figure 16. For detected systems (brighter than ~25mag) the average radio source radii are in the range 0.6 – 0.8 arcseconds implying that the next generation radio interferometers will need sub-arcsec angular resolution.

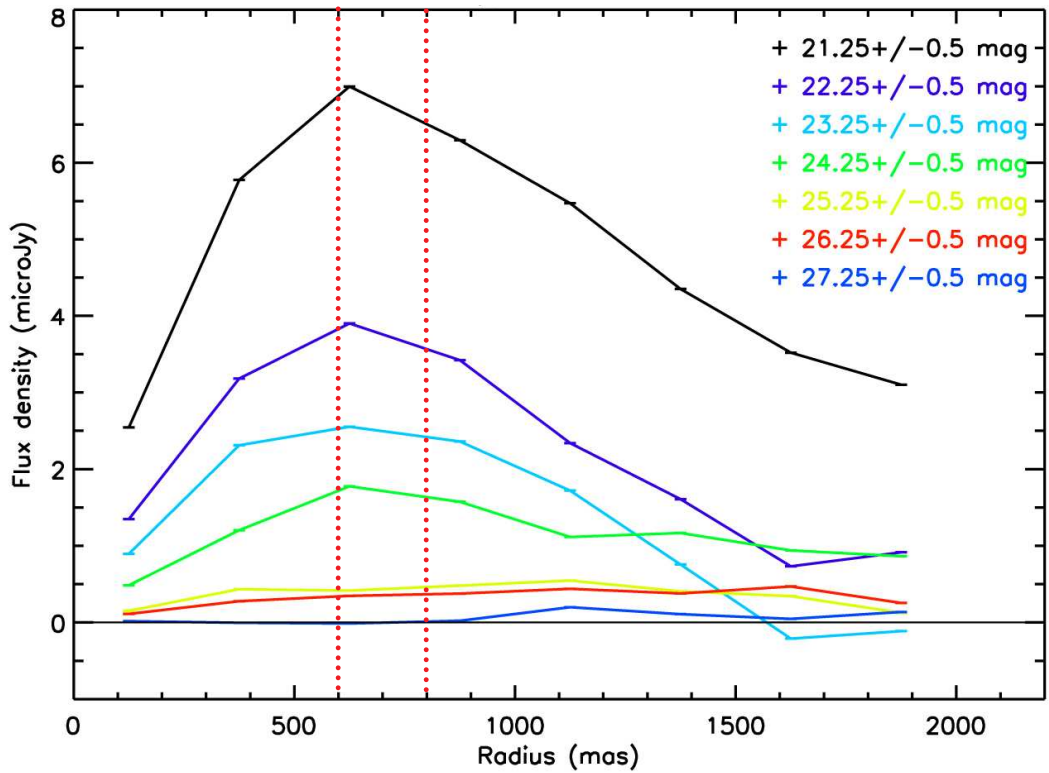

*Figure 16: The average radio flux density contained within annuli of width 0.5" over radii of 0.25" to 2" plotted for each Z-band magnitude bin. For galaxies brighter than Z~25$^{mag}$ the average radio source radii are in the range 0.6" to 0.8".*

Source sizes may also be measured by fitting to the composite, stacked radio image in each magnitude bin. Similar sizes are derived by this latter method. Figure 17 shows the averaged composite optical and radio images for the 23$^{mag}$ Z-band bin. This contains some 927 galaxies and it is interesting to note that the stacked radio image has an rms noise level of 108nJy/beam, which is a factor of ~30 lower than that in the original radio image and in line with expectations for averaging ~900 regions of the original image.

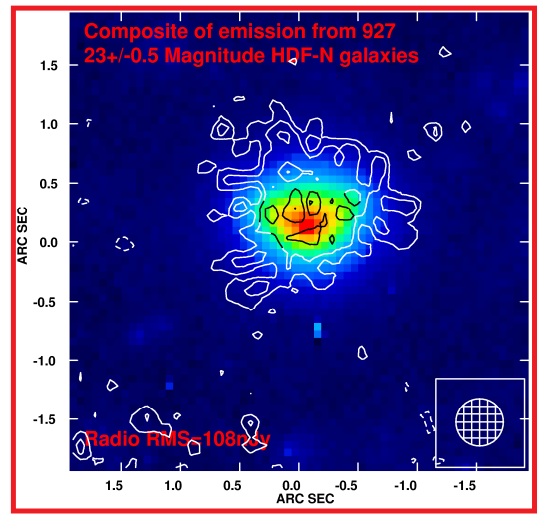

*Figure 17: Composite optical (false colour) and radio image (contoured) for the Z-band 23$^{mag}$ bin. Contours are linear with an interval of 3σ (σ=108nJy/beam).*

### 4.4 Starburst Luminosities

Only ~1000 of the ~13000 galaxies in the 8.5' field have published spectroscopic redshifts available. We have constructed the luminosity-redshift distribution for all those starburst systems with measured redshifts. For this sub-set of starbursts, the detected radio flux density stacked by Z-band magnitude is not significantly different from the complete galaxy sample (at least for those magnitude bins brighter than $24^{mag}$); indicating that the sub-set of galaxies with measured redshifts do not appear to differ significantly from the complete galaxy set. We are thus able to derive the averaged luminosity distribution for those starburst galaxies brighter than $24^{th}$ magnitude. This is shown in figure 18 together with the individual starburst systems with 1.4GHz flux densities >40 µJy taken from [24].

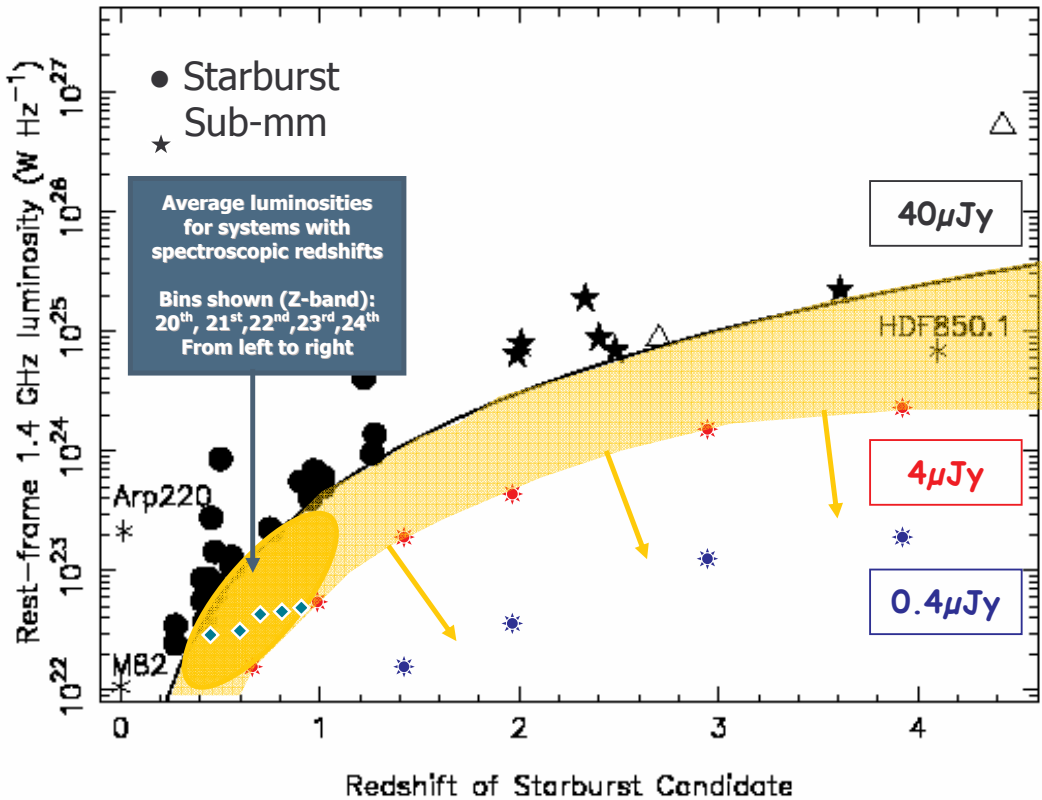

*Figure 18: Rest frame monochromatic luminosities for the individual starburst galaxies taken from [23] and the averaged luminosities for Z-band magnitude bins. 40µJy sample cut-off is shown as a line. 4µJy and 0.4µJy cut-off shown as coloured symbols. Individual starburst systems from [24] are shown as filled circles. Sub-mm detections are shown as filled stars. Two high redshift starburst thought to contain embedded AGN are shown as open triangles. The statistical results for the very weak radio sources are shown as filled blue diamonds. For comparison two nearby starbursts (M82 & Arp220) are also shown together with HDF850.1, the brightest sub-mm SCUBA detection in the HDF-N.*

The very weak radio sources which are assumed to be dominated by starbursts have average properties which indicate that they are an extension of the weaker end of the distribution of starbursts detected as individual sources at higher radio flux densities. However, it should be noted that these weak radio sources have been selected by optical Z-band properties and will therefore exclude optically faint systems. Most of the individual high redshift sub-mm starburst systems imaged by [24] were optically faint, and therefore will not be represented statistically in the weak source results. Clearly individual imaging of these very faint radio sources will be required. The new upgraded radio interferometers due to be commissioned in the next few years will be able to achieve this. *e*-MERLIN, EVLA and *e*-VLBI should be able to image >1000 individual starburst systems to flux densities of ~4μJy at 1.4GHz with perhaps 150-200 objects at high redshift in a single field.

For the even weaker radio source population, the new instruments will be able to extend this present statistical study to many thousands of starburst systems with radio flux densities <1μJy. SKA and ALMA will ultimately extend this by an additional order of magnitude. With more redshifts, improved SED templates, and extinction-free SFR indicators it will be possible to solve for the cosmic star-formation history – The Radio Madau diagram

### 4.5 Star-formation History

As an example of work in progress, we can report an initial investigation into the cosmic star-formation history by utilizing the Spitzer IR detections in this field. Within the 8.5 arcminute field there are 377 Spitzer 24μm detections complete to 80μJy [27]. Of these, 303 (~80%) are detected in our combination radio image (integrated flux density >3σ within 0.75 arcseconds). Restricting the sample to those detected in the radio which also have measured redshifts yields a final sample of 213 star-forming galaxies. We have used these to investigate the evolution of star-formation density with redshift. The results are shown in figure 19. At present, we can show that the SFR density increases dramatically to z~1 and then flattens [28]. At higher redshifts, the position of the turnover point remains uncertain and requires new data from the next generation radio interferometers which will image the sub-μJy radio source population.

The next few years will be very exciting…

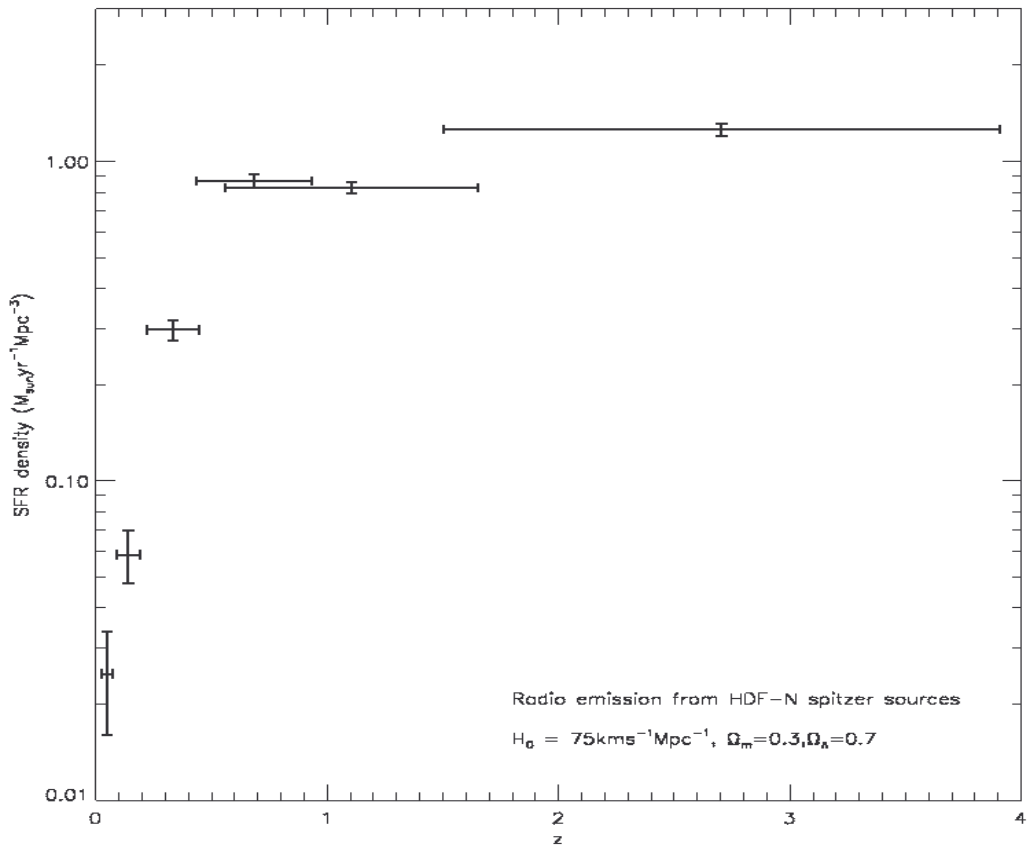

*Figure 19: Star-formation density – redshift distribution derived from the radio-detected Spitzer sources in the HDF-N with measured redshifts. Interim results – Beswick (Private Communication)*